# Light tunneling inhibition in longitudinally modulated Bragg-guiding arrays

Yaroslav V. Kartashov and Victor A. Vysloukh

*ICFO-Institut de Ciencies Fotoniques, and Universitat Politecnica de Catalunya, Mediterranean Technology Park, 08860 Castelldefels (Barcelona), Spain*

We consider waveguide array in which defect channels with reduced refractive index are spaced periodically and the light guiding is achieved due to Bragg reflection. We show that tunneling between defects can be inhibited using out-of-phase longitudinal modulation of refractive index exclusively in defect channels. Tunneling inhibition is almost perfect for strongly localized modes with propagation constants close to the gap center.

The materials with inhomogeneous refractive index landscapes offer exceptional opportunities for diffraction control. A guiding based on Bragg reflections (BR) is possible in periodic structures in the presence of nonlinearity or defects even when they are negative in contrast with guiding due to total internal reflection (TIR) occurring around global refractive index maxima [1,2]. The longitudinal refractive index modulation enriches the opportunities for control of light evolution and allows the formation of diffraction-managed solitons [3,4] and dynamic localization of light [5-19]. It was studied in driven double-well potentials [5,6], optical couplers [7-9], and waveguide arrays [10-19], where periodic bending [10,13,15] or out-of-phase modulation of refractive index in adjacent guides [11,12,18,19] can be used for light tunneling control. The inhibition of tunneling is a resonant effect with resonant frequencies depending on modulation laws [20,21]. However, all these effects were demonstrated in arrays where light trapping in guides was due to TIR. The question whether tunneling inhibition is possible when guiding occurs due to BR remains open. Notice, that even in geometric-optics approach the guiding due to TIR is essentially different from that due to BR at least because TIR occurs in the single band of spatial frequencies while BR appears for a sequence of Bragg resonances. Floquet-Bloch formalism predicts that not only profiles of nonlinear modes, but also domains of their existence depend crucially on whether the mode is supported by TIR or BR [1,2].

In this Letter we consider light propagation in waveguide arrays with periodically spaced defect channels with reduced refractive index and show that in such structure the light tunneling can be inhibited due to out-of-phase refractive index modulation in defect channels. A consequence of BR guiding is that the modulation frequencies at which inhibition can be achieved and the degree of tunneling inhibition are nonmonotonic functions of refractive index in defect channels.

The propagation of light along the $\xi$-axis of array is described by the nonlinear Schrödinger equation for the dimensionless field amplitude $q$:

$$i\frac{\partial q}{\partial \xi} = -\frac{1}{2}\frac{\partial^2 q}{\partial \eta^2} - |q|^2 q - R(\eta,\xi)q. \tag{1}$$

Here $\eta$ and $\xi$ are the normalized coordinates, $R(\eta,\xi)$ describes refractive index profile in array of guides $G(\eta) = p(\xi)\exp(-\eta^6/a^6)$ with refractive index contrast $p(\xi) \sim \delta n = n_{\text{pert}} - n_0$ ($n_{\text{pert}}, n_0$ are perturbed and unperturbed refractive indices), width $a$, and spacing $d$. In the absence of longitudinal modulation the refractive index $p = p_d$ in defect guides that are periodically nested into array and are separated by $m$ usual guides with $p_a > p_d$. In the presence of modulation refractive index in defect changes as $p_d \pm \mu\sin(\Omega\xi)$ ($\mu$ and $\Omega$ are the amplitude and frequency of modulation; oscillations in neighboring defects are out-of-phase), but it is constant in usual guides. Further we set $a = 0.3$ and $d = 1.4$; $p \simeq 7$ is equivalent to $\delta n \simeq 7.8 \times 10^{-4}$ at $\lambda = 800$ nm [18].

In the absence of defects $(p_d = p_a)$ the linear modes are Bloch waves $q = w(\eta)\exp(ib\xi + i\kappa\eta)$ with $w(\eta) = w(\eta + d)$. The propagation constants $b$ form bands separated by the gaps [Fig. 1(a)]. Localized modes appear when $p_d \neq p_a$ at least in one channel. If $p_d < p_a$ the linear modes form due to BR, they have oscillating tails [Figs. 1(b) and 1(d)], and $b$ values in the gaps. The mode width depends on the position of $b$ in the gap [Fig. 1(e)]. While modes with $b$ values in the deep of the gap are well localized [Fig. 1(c)], their counterparts with $b$ values close to gap edges are extended [Figs. 1(b) and 1(d)]. When $p_d \to p_a$ the $b$ value approaches the upper gap edge, while with decrease of $p_d$ it approaches the lower gap edge. If there are two or more defects, one can observe tunneling of light between them [see Fig. 4(b)]. The tunneling scale is characterized by the period $T_b = 2\pi/\Omega_b$ of energy beating in a pair of defects. At $p_d < p_a = 7$ the period $T_b$ (hence coupling constant) is a nonmonotonic function of $p_d$ reaching its maximum in the deep of the gap [Fig. 1(f)]. When $p_d > p_a = 7$ and guiding is due to TIR, $T_b$ monotonically grows with $p_d$.

The effect of tunneling inhibition in linear case can be understood using tight-binding approximation that operates with fields $q_m$ in $m$-th defect guide that satisfy the equation $idq_m/d\xi + (-1)^m \mu \sin(\Omega\xi) q_m + C(q_{m+1} + q_{m-1}) = 0$, with $C$ being the coupling constant. The transformation $h_m = q_m \exp[i(-1)^m \mu \cos(\Omega\xi)/\Omega]$ yields $idh_m/d\xi + C \exp[2i\mu(-1)^m \cos(\Omega\xi)/\Omega](h_{m+1} + h_{m-1}) = 0$. Direct integration of this system shows that distance-averaged power in the excited channel $U_m = L^{-1} \int_0^L |h_0(\xi)|^2 d\xi$ ($L$ is the propagation distance) attains maximal values for a sequence of resonant frequencies $\Omega$. Remarkably, for $\mu/C \gg 1$ [$\mu$ value here is not to be mixed with $\mu$ value in Eq. (1)] the $U_m(\Omega)$ dependence in the model with modulated coupling constant $C \exp[2i\mu(-1)^m \cos(\Omega\xi)/\Omega]$ is very close to dependence obtained in the model where this constant is replaced with $CJ_0(2\mu/\Omega)$ where $J_0(2\mu/\Omega)$ is first term in expansion $\exp[2i\mu \cos(\Omega\xi)/\Omega] = \sum_k i^k J_k(2\mu/\Omega) \exp(ik\Omega\xi)$. This is expected since impact of the term with $k = 0$ on system dynamics is more pronounced than impact of terms with $|k| > 0$ containing oscillating factors $\sim \exp(ik\Omega\xi)$. Thus longitudinal modulation causes effective modulation of coupling constant and may result in considerable inhibition of tunneling when effective coupling is minimal. While tight-binding description is similar for BG and TIR cases, the dependencies $C(p_d)$ for these two mechanisms are remarkably different [compare curves in Fig. 1(f) for $p_d < p_a$ and $p_d > p_a$ showing $T_b(p_d) \sim 1/C$].

The important result of this Letter is that longitudinal out-of-phase refractive index modulation in neighboring defects does allow tunneling inhibition as confirmed by simulations of Eq. (1). To achieve tunneling inhibition it is sufficient to modulate refractive index only inside the defect channels. This is different from tunneling inhibition in usual arrays, where all waveguides should be modulated out-of-phase [18]. To quantify tunneling inhibition, we introduce the distance-averaged power in the excited (central) defect: $U_m = L^{-1} \int_0^L d\xi \int_{-d/2}^{d/2} |q(\eta,\xi)|^2 d\eta \Big/ \int_{-d/2}^{d/2} |q(\eta,0)|^2 d\eta$, where $L = 4T_b$. We used $q|_{\xi=0} = Aw(\eta)$ as an initial condition, where $w(\eta)$ is the profile of defect mode and $A$ is its amplitude. Typical dependence $U_m(\Omega)$ is shown in Fig. 2(a) in the linear limit ($A \to 0$) for $m = 2$. Tunneling inhibition is a resonant effect, which is possible for strictly defined set of resonance frequencies (further we will normalize $\Omega$ to the beating frequency $\Omega_b$). There exists principal resonance with $\Omega = \Omega_r$, while frequencies of secondary resonances $\approx \Omega_r/k$, with $k = 2, 3, \ldots$. Usually, principal resonance is the strongest one and corresponds to maximal $U_m(\Omega_r)$ value (we denote it $U_{max}$). For intermediate defect strengths $U_{max}$ increases with $\mu$, but only until $\mu \leq 0.15 p_a$; $\Omega_r$ grows almost linearly with $\mu$ [Fig. 2(b)]. The nonlinearity facilitates inhibition - increasing $A$ causes substantial broadening of resonances in Fig. 2(a).

The width $\delta\Omega/\Omega_{\rm b}$ of primary resonance (at the level $0.7U_{\max}$) increases with $A$ and for sufficiently high $A$ values $\delta\Omega/\Omega_{\rm b} \sim A^2$ [Fig. 2(c)].

Figure 3 illustrates the dependence of $\Omega_{\rm r}$ and $U_{\max}$ on $p_{\rm d}$. Variation of $p_{\rm d}$ results in dramatic changes of the guided mode width and the energy tunneling rate. Moreover, modifications in $p_{\rm d}$ noticeably affect $\Omega_{\rm r}$ that acquires its maximal value for certain $p_{\rm d}$ at which mode of isolated defect is strongly localized [Fig. 3(a)]. $\Omega_{\rm r}$ decreases considerably when $p_{\rm d} \to p_{\rm a}$ and if $p_{\rm d}$ takes sufficiently large negative values. Tunneling inhibition is almost perfect ($U_{\max} \simeq 1$) for $p_{\rm d}$ values around which $\Omega_{\rm r}$ is maximal [Fig. 3(b)]. Deviation of $p_{\rm d}$ from the optimal value results in substantial reduction of $U_{\max}$. This feature becomes clear in the limit $p_{\rm d} \to p_{\rm a}$, when defect modes are weakly localized and longitudinal modulation which lifts defect's refractive index up to the value $p_{\rm d} + \mu$ may result in dramatic radiation. In this limit $U_{\max}$ may decrease with $\mu$ in contrast to the scenario encountered for $p_{\rm d}$ values around which $\Omega_{\rm r}/\Omega_{\rm b}$ has a maximum. If $p_{\rm d}$ takes considerable negative values, one should increase $\mu$ to observe at least primary resonance in $U_{\rm m}(\Omega)$ dependence. Poor inhibition of tunneling can be attributed to weak localization of defect modes. An example of refractive index distribution is depicted in Fig. 4(a), while diffraction in the absence of longitudinal modulation is shown in Fig. 4(b). Longitudinal modulation dramatically slows down the rate of light spreading [Fig. 4(c)]. An almost perfect tunneling inhibition is illustrated in Fig. 4(d) at $\Omega = \Omega_{\rm r}$. Figures 4(e) and 4(f) show how the quality of tunneling inhibition in primary resonance diminishes when $p_{\rm d}$ shifts from the optimal value at fixed $\mu$.

Summarizing, we showed that tunneling inhibition is possible in waveguide arrays with periodically nested negative defects. We found that tunneling inhibition is most pronounced when maximal localization of defect modes is achieved and frequency of primary resonance takes maximal value. It is the guiding mechanism which makes the physics of tunneling inhibition in BG arrays substantially richer than in conventional periodic arrays.

# References with titles

# Figure captions

Figure 1.  (a) Bandgap spectrum of uniform array. Dashed line corresponds to $p_a = 7$. (b)-(d) Profiles of defect modes at $p_a = 7$. Gray regions - usual guides, cyan region - defect guide. (e) $b$ versus $p_d$ for defect modes. Points correspond to modes in (b)-(d). (f) $T_b$ versus $p_d$ for two defects separated by two usual guides at $p_a = 7$.

Figure 2.  (a) $U_m$ versus $\Omega$ at $\mu = 0.5$. Dashed line corresponds to $U_m = 1$. (b) $\Omega_r$ versus $\mu$. (c) $\delta\Omega$ as a function of $A^2$ at $\mu = 0.8$. In all cases $p_d = 4$, $p_a = 7$.

Figure 3.  (a) $\Omega_r$ and (b) $U_{\max}$ versus $p_d$ at $\mu = 0.8$, $p_a = 7$. Dashed line in (b) corresponds to $U_m = 1$.

Figure 4.  (a) An example of $\xi$-modulated array of Bragg guides. The refractive index at $\xi = 0$ is shown in the bottom of the panel. (b) Discrete diffraction in unmodulated array with $p_d = 4.0$. Propagation dynamics in modulated arrays at (c) $p_d = 4$, $\Omega = 5.70\Omega_b$, (d) $p_d = 4$, $\Omega = 7.47\Omega_b$, (e) $p_d = 1$, $\Omega = 3.12\Omega_b$, and (f) $p_d = 5.8$, $\Omega = 5.12\Omega_b$. In (c)-(f) $\mu = 0.8$, while propagation distance is $\xi = 200$. In all cases $A = 0.1$, $p_a = 7$.

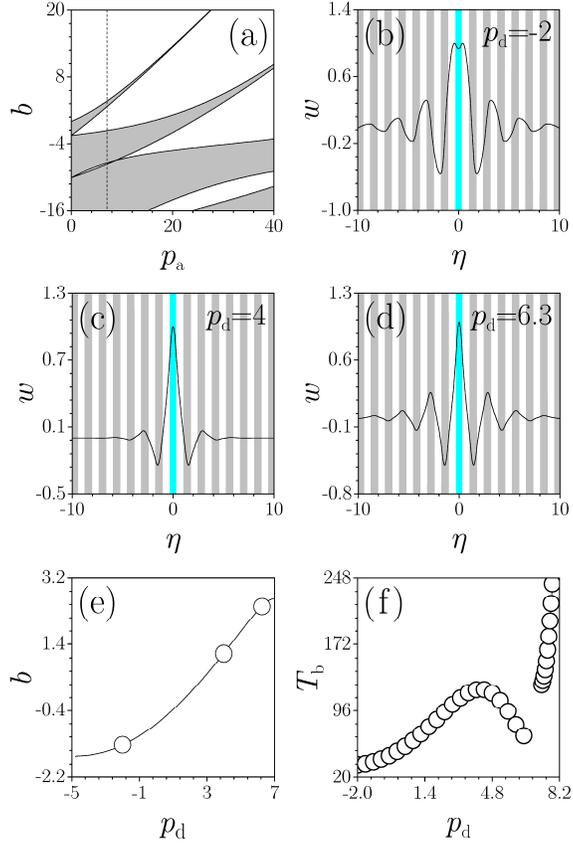

Figure 1.  (a) Bandgap spectrum of uniform array. Dashed line corresponds to $p_\mathrm{a}=7$. (b)-(d) Profiles of defect modes at $p_\mathrm{a}=7$. Gray regions - usual guides, cyan region - defect guide. (e) $b$ versus $p_\mathrm{d}$ for defect modes. Points correspond to modes in (b)-(d). (f) $T_\mathrm{b}$ versus $p_\mathrm{d}$ for two defects separated by two usual guides at $p_\mathrm{a}=7$.

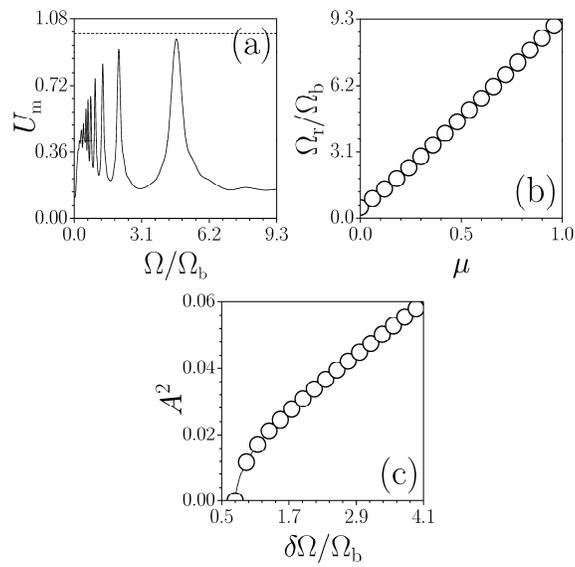

Figure 2. (a) $U_\mathrm{m}$ versus $\Omega$ at $\mu = 0.5$. Dashed line corresponds to $U_\mathrm{m} = 1$. (b) $\Omega_\mathrm{r}$ versus $\mu$. (c) $\delta\Omega$ as a function of $A^2$ at $\mu = 0.8$. In all cases $p_\mathrm{d} = 4$, $p_\mathrm{a} = 7$.

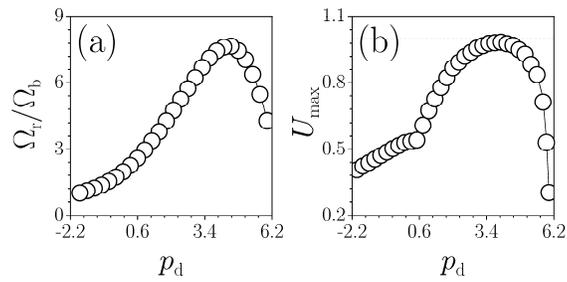

Figure 3. (a) $\Omega_r$ and (b) $U_{\max}$ versus $p_d$ at $\mu = 0.8$, $p_a = 7$. Dashed line in (b) corresponds to $U_m = 1$.

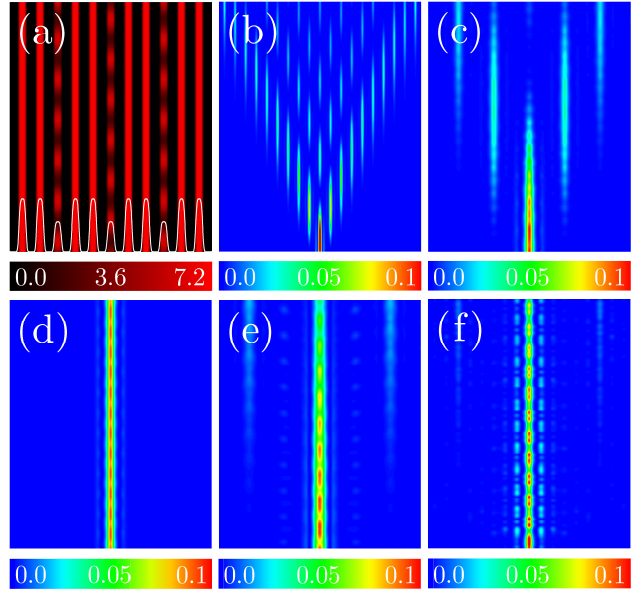

Figure 4. (a) An example of $\xi$-modulated array of Bragg guides. The refractive index at $\xi=0$ is shown in the bottom of the panel. (b) Discrete diffraction in unmodulated array with $p_{\rm d}=4.0$. Propagation dynamics in modulated arrays at (c) $p_{\rm d}=4$, $\Omega=5.70\Omega_{\rm b}$, (d) $p_{\rm d}=4$, $\Omega=7.47\Omega_{\rm b}$, (e) $p_{\rm d}=1$, $\Omega=3.12\Omega_{\rm b}$, and (f) $p_{\rm d}=5.8$, $\Omega=5.12\Omega_{\rm b}$. In (c)-(f) $\mu=0.8$, while propagation distance is $\xi=200$. In all cases $A=0.1$, $p_{\rm a}=7$.